\documentclass[fleqn,usenatbib]{mnras}

\usepackage{newtxtext,newtxmath}
\usepackage{threeparttablex}

\usepackage[T1]{fontenc}

\DeclareRobustCommand{\VAN}[3]{#2}
\let\VANthebibliography\thebibliography
\def\thebibliography{\DeclareRobustCommand{\VAN}[3]{##3}\VANthebibliography}

\usepackage{graphicx}	
\usepackage{amsmath}
\usepackage{subfigure} 

\usepackage[normalem]{ulem}  
\newcommand\redout{\bgroup\markoverwith
{\textcolor{red}{\rule[0.5ex]{2pt}{0.8pt}}}\ULon}

\newcommand{\ac}[1]{\textcolor{orange}{#1}} 

\usepackage{tikz,xcolor,hyperref}

\title[Rectangular and jet-like SNR morphologies]{The sculpting of rectangular and jet-like morphologies in supernova remnants by anisotropic equatorially-confined progenitor stellar winds}

\author[P. F. Velázquez et al.]
{P. F. Velázquez,$^{1}$\thanks{E-mail: pablo@nucleares.unam.mx (PFV)}
D. M.-A. Meyer,$^{2}$
A. Chiotellis,$^{3,4}$
A. E. Cruz-Álvarez,$^{1}$
E. M. Schneiter,$^{5}$\and
J. C. Toledo-Roy,$^{1}$
E. M. Reynoso,$^{6}$
and A. Esquivel$^{1}$
\\
$^{1}$ Instituto de Ciencias Nucleares, Universidad Nacional Aut\'onoma de M\'exico, Ap. 70-543, CDMX, 04510, M\'exico\\
$^{2}$ Universit\" at Potsdam, Institut f\" ur Physik und Astronomie, 
                 Karl-Liebknecht-Strasse 24/25, 14476 Potsdam, Germany\\
$^{3}$ Institute for Astronomy, Astrophysics, Space Applications and Remote Sensing, 
                 National Observatory of Athens, 15236, Penteli, Greece \\
$^{4}$ 4rth Lykeion Acharnon, Acharneon Ippeon and Paliggenesias, 136 74   Acharnes, Greece\\
$^{5}$ Departamento de Materiales y Tecnolog\'{i}a, FCEFyN-UNC, Av. V\'elez Sarsfield 1611, C\'ordoba, Argentina \\ 
$^{6}$ Instituto de Astronom\'{i}a y F\'{i}sica del Espacio (IAFE), Av. Int. G\"uiraldes 2620, 
               Pabellón IAFE, Ciudad Universitaria, 1428, Buenos Aires, Argentina
}

\date{Accepted XXX. Received YYY; in original form ZZZ}

\pubyear{2015}

\begin{document}
\label{firstpage}
\pagerange{\pageref{firstpage}--\pageref{lastpage}}
\maketitle

\begin{abstract}
Thermonuclear and core-collapse supernova remnants (SNRs) are the nebular leftovers of defunct stars. Their morphology and emission properties provide insights into the evolutionary history of the progenitor star. But while some SNRs are spherical, as expected from a point-like explosion expanding into a roughly uniform medium, many others exhibit complex non-spherical morphologies which are often not easily explained. In this work, we use three-dimensional magnetohydrodynamic simulations to show that rectangular and jet-like morphologies can be explained by supernovae (SNe), either type Ia or type II, expanding within anisotropic, bipolar stellar wind bubbles driven by the progenitor star.
The stellar wind has an anisotropic density distribution, which channels the SN ejecta differently depending on the anisotropy characteristics. We compute synthetic thermal (X-ray) and non-thermal (synchrotron) emission maps from our numerical simulations to compare with observations. We find rectangular morphologies are generated when the stellar wind has a high mass loss rate and forms a dense, narrow disk at the equatorial region. Instead, a jet-like or ear-like morphology is obtained when the stellar wind develops a wide, dense disk. Stellar winds with low mass-loss rates do not strongly influence the SNR morphology. Finally,  our synthetic synchrotron and X-ray maps for the high mass-loss rate case qualitatively agree with the observations of the SNRs G332.5-5.6 and G290.1-0.8.

\end{abstract}




\section{Introduction}\label{intro}
\begin{figure*}
    \centering
    \includegraphics[width=14cm]{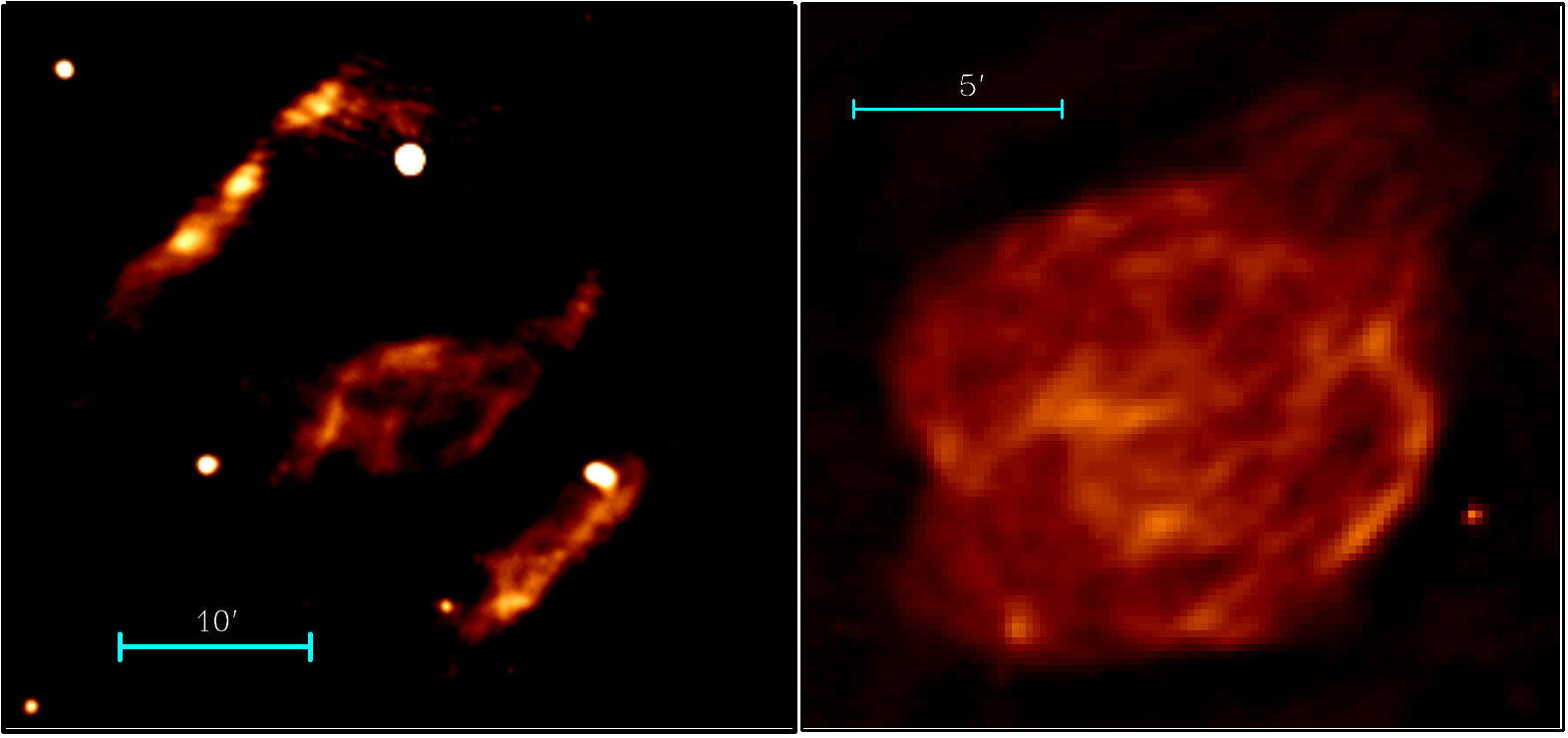}
    \caption{Radiocontinuum observations at 1.4 GHz of SNRs G332.5-5.6 \citep[left,][]{reynoso2007} and G290.1-0.8 \citep[right,][]{reynoso2006}. The blue line indicates the angular-size in arc-minutes}
    \label{fig:SNRs}
\end{figure*}

A supernova (SN) explosion is one of the most energetic events in the Universe, depositing in a few seconds $10^{51}$~erg of energy and a few solar masses of ejecta into the surrounding interstellar medium (ISM) at velocities in excess of $\simeq 10^4$ km s$^{-1}$. This phenomenon produces strong shock waves that heat up and compress the surrounding ISM and the freely expanding ejecta. The region enclosed by these shock waves is known as supernova remnant (SNR), containing the shocked and freely expanding ejected material as well as the swept-up ISM.

Depending on the explosion's physical mechanism supernovae (SNe) are divided into two major classes: those originating from the thermonuclear explosion of a carbon-oxygen white dwarf (WD) in an interacting binary system and those resulting from the gravitational collapse of the core  of a massive star \citep{Vink2020}.  In both cases, the ambient medium is very likely to be modified by the strong stellar winds blown during the evolution of massive stars -for the case of core-collapse SNe \citep[e.g.][]{Chita2008,meyer_mnras_450_2015, meyer_mnras_493_2020, Yamaguchi2021,Boumis2022,herbst_2022}, or by the mass outflows emanating from the WD’s surface and/or from its companion star in the case of thermonuclear SNe \citep[e.g.][]{Chiotellis2012, Chiotellis2013, Toledo-Roy2014, Broersen2014}.  Subsequently, when the SN explosion occurs, the stellar ejecta interacts with the previously modified ambient medium and the effects of this interaction will be reflected on the morphology of the resulting SNR. Hence, the study of the vast variety of SNR shapes and morphologies provide us with key insights into the properties of both the ambient medium and the SN progenitor \citep[e.g.][]{Lopez2011,das_2022}, as well as into the SN explosion mechanism itself \citep{woosley_araa_24_1986, weiler_araa_25_1988, filippenko_araa_35_1997, Badenes2006, smartt_araa_47_2009, Yamaguchi2014,2022arXiv220804875S}.

Some SNRs present roughly spherical morphologies but in addition have two symmetrically opposite ear-like protrusions.
Some examples in this group are the SNRs W50 \citep{dubner1998}, G309.2-0.6 \citep{gaensler1998}, and G290.1-0.8 \citep{reynoso2006}, the latter shown in the right panel of Figure \ref{fig:SNRs} \citep[see][for more examples]{tsebrenko15a}. At the same time, other SNRs exhibit quadrilateral shapes, as for example Puppis A \citep{reynoso2017,reynoso2018} and G332.5-5.6 \citep[left panel of Figure \ref{fig:SNRs}; e.g. ][]{reynoso2007}. 

Several works have addressed the origin of ear-like SNRs, proposing different scenarios. Some of these works suggest that two narrow and opposite jets are responsible for this observed morphology and can appear before or during the SN explosion. These jet-like structures could be produced by the accretion of matter by a white dwarf star in a type Ia SN, or the neutron star formation in a core-collapse SN \citep[][]{velazquez2000,zavala2008, soker2010,tsebrenko2013,papish2014,akashi2021}. Another scenario assumes the presence of inhomogeneities in the density and expansion velocity distribution of the ejecta, as proposed in \citet{tsebrenko2015c}, \citet{orlando2016} and \citet{tutone2020}.

An alternative idea is that the medium surrounding the progenitor  is responsible for these non-spherical morphologies. These models assume that the SNR expansion takes place in an axisymmetric, but not spherically symmetric, circumstellar medium (CSM) formed by the progenitor stellar wind, either in the Asymptotic Giant Branch \citep[AGB;][]{2022arXiv220614806S} or the Red Super Giant (RSG) 
phase \citep{meyer_mnras_506_2021}. The axisymmetry results from a density increase of the AGB or RSG winds towards the equator. \citet{blondin1996} explored this scenario, assuming a type II SN explosion and an axisymmetric CSM formed by an RSG stellar wind. Later, \citet{tsebrenko15b} modelled the SNR G1.9+0.3, considering that the remnant evolves inside the progenitor's planetary nebula. More recently, \citet{ustamujic2021}, also imposing an axisymmetric CSM, studied the case where the SNR's progenitor is a luminous blue variable (LBV) star. In all cases, the ear-like or jet-like features are found to be produced along the polar axis (the symmetry axis) of the wind. However, in a recent work, \citet{alexandros21} explored a similar scenario including density and velocity dependencies on the polar angle in the progenitor's stellar wind, and found that ear-like features can also be formed in the equatorial regions of the remnant.

Recently, \citet{2022arXiv220614495M} studied the formation of SNRs with rectangular morphologies from a massive progenitor (35~M$_\odot$). The remnant interacts with the stellar bubble resulting from the complex interaction of multiple subsequent wind phases: the O-type (main sequence) wind, the RSG phase, and finally a Wolf-Rayet (WR) wind. This evolution takes place in an ISM  with a uniform magnetic field, producing an asymmetric circumstellar medium in which the SN ejecta expands, and results in a projected rectangular morphology. Other studies on (a)symmetric core-collapse SNRs include \citet{meyer_mnras_450_2015,meyer_mnras_493_2020,Meyer2021}, \citet{orlando_aa_645_2021,2022arXiv220201643O}, \citet{soker_2021}, and \citet{soker_2021_b}. 

In this work, we further explore the idea of an anisotropic stellar wind proposed by \citet{blondin1996} and \citet{alexandros21}. By considering several azimuthal distributions of the stellar wind, we explore the final morphologies of the resulting SNRs and study whether such a scenario can reproduce and qualitatively explain the formation of  SNRs with jet-like and rectangular morphologies. Unlike \citet{2022arXiv220614495M}, in this work, we will focus on studying SNRs from progenitors with stellar masses $\leq 16\, \mathrm{M}_\odot$ in the main sequence stage.

We organise the present work as follows:  Section \ref{Sec:2} describes the characteristics of the studied scenarios and the initial setup of the simulations. The results of our modelling and their analysis are presented in Section \ref{Sec:3}. Finally, we summarise our main conclusions in Section \ref{Sec:4}.

\section{Simulations: initial setup}\label{Sec:2}
\subsection{The scenario}

\begin{figure}
    \centering
    \subfigure[]{\includegraphics[width=\columnwidth]{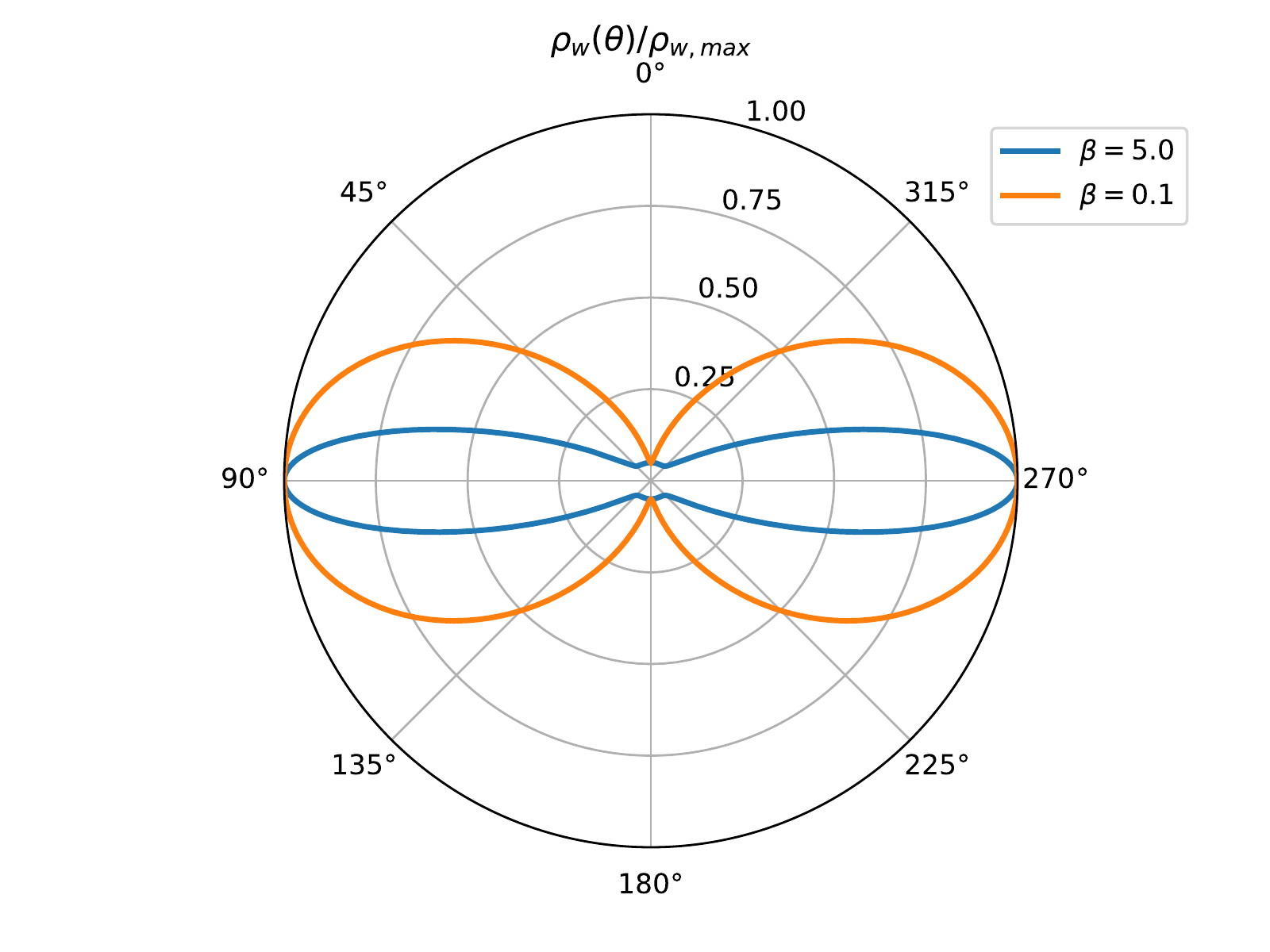} \label{fig:rhow}}
    \subfigure[]{	\includegraphics[width=\columnwidth]{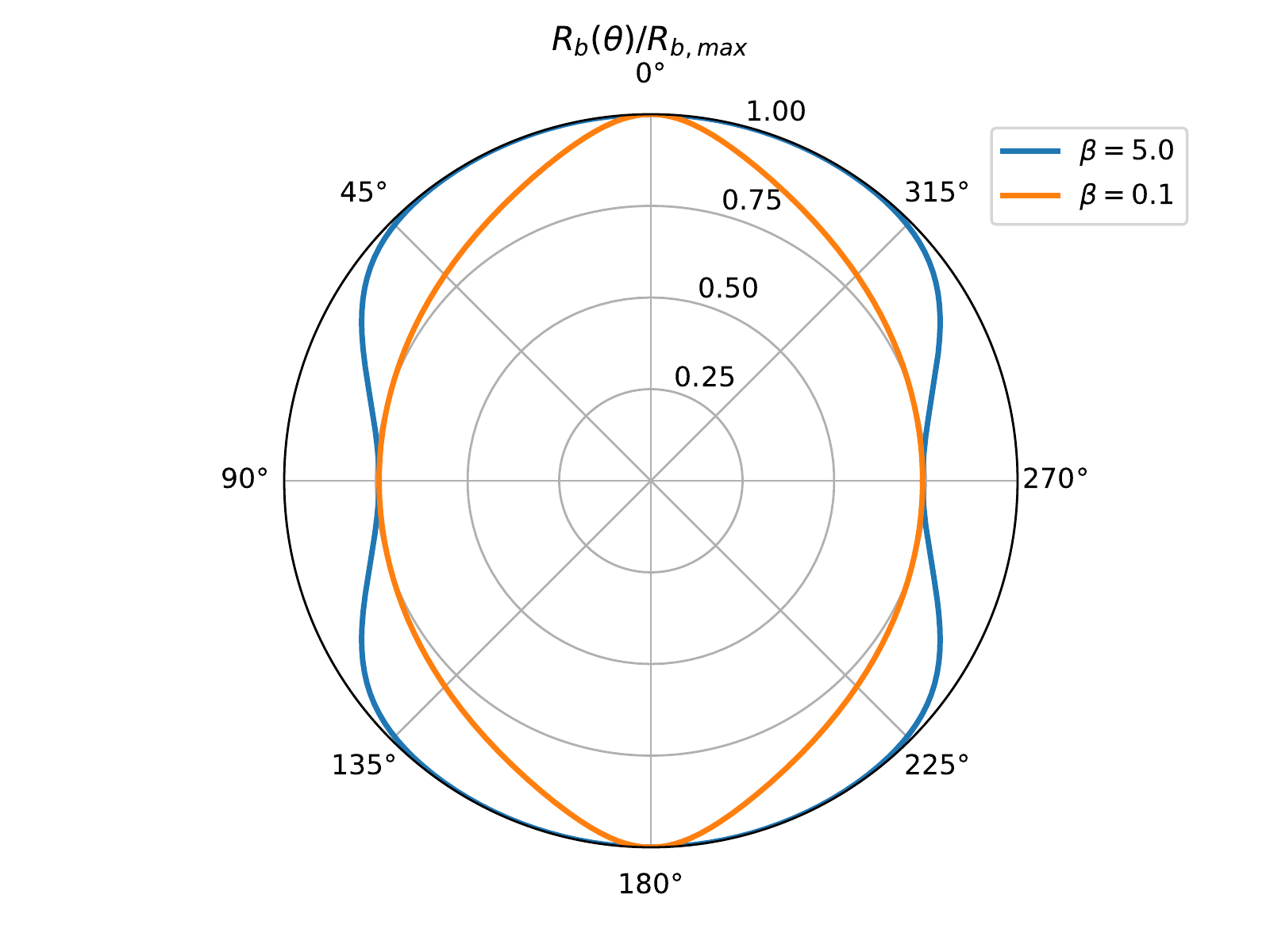}\label{fig:rbw}}
    \caption{Polar dependence of (a): the normalised density distribution $\rho_{\rm w}(\theta)/\rho_\textrm{w,max}$   and (b): the wind bubble outer radius $R_{\rm b}(\theta)/R_\textrm{b,max}$  for a narrow ($\beta=5.0$) and wide  ($\beta=0.1$) equatorial disk.}
    \label{fig:rbw_rhow}
\end{figure}



\begin{figure*}
    \centering
    \includegraphics[width=15cm]{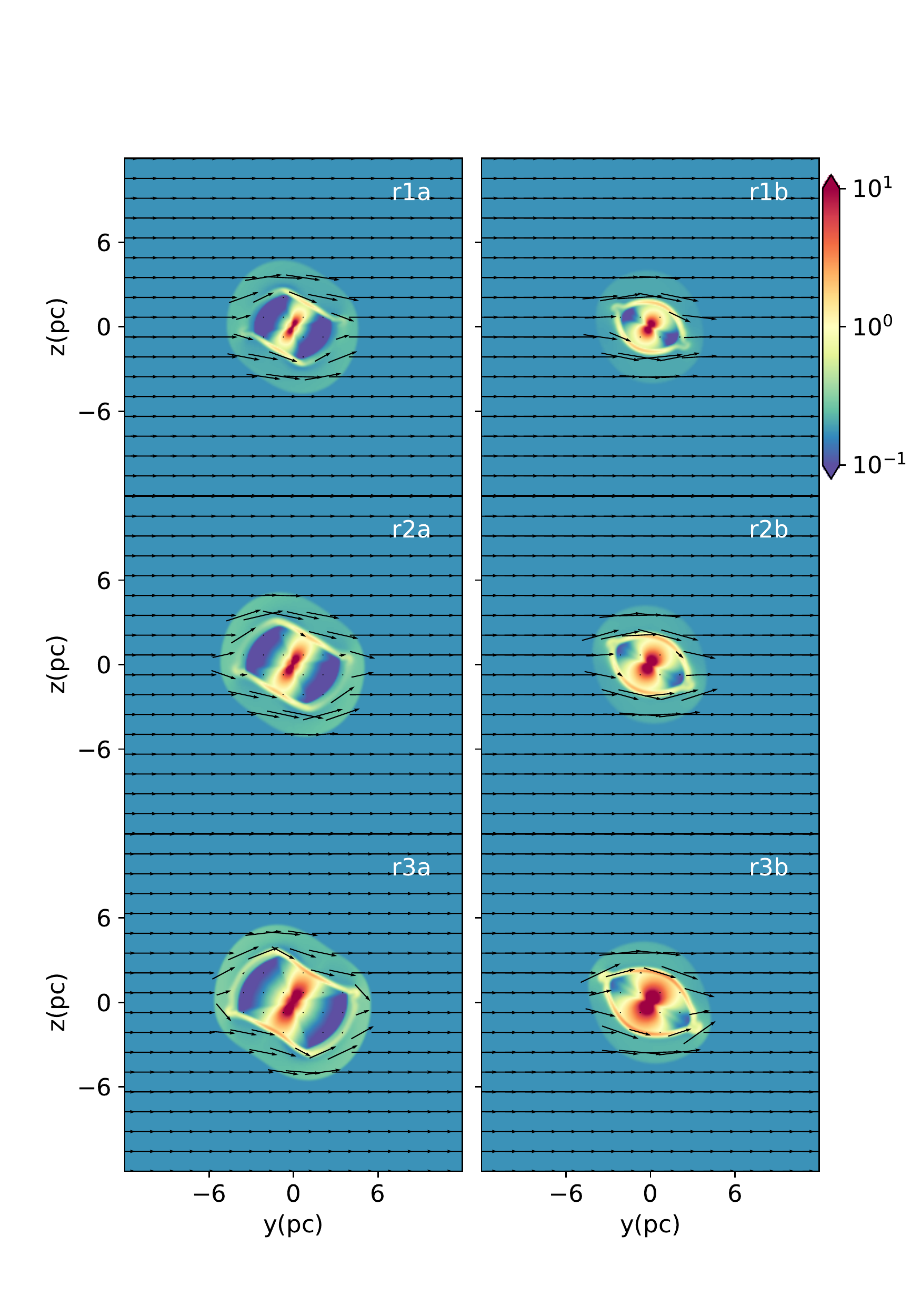}
    \caption{Density stratification on the $yz-$ plane of the stellar wind bubble obtained for all runs,  which are indicated by the label on the top-right corner on each map. These maps show an integration time of 300~kyr, corresponding to the moment just before the SN explosion. The density is given in units of $2.5\times10^{-24}\,\mathrm{g\, cm^{-3}}$. The black arrows indicate the direction of ISM magnetic field. 
    The horizontal and vertical axes units are in pc.
    }
    \label{fig:winds}
\end{figure*}

Our goal is to propose a general model which can explain both the box-like and the jet-like morphologies observed in a number of SNRs. In Section \ref{intro}, we mentioned a number of these peculiar astrophysical objects, with Puppis A and G332.5-5.6 being examples of the first group of SNRs, and G309.2-0.6 and G290.1-0.8 belonging to the latter group. These sources have physical radii  ranging from 9 to 15 pc. Therefore, we choose a Cartesian computational domain spanning 24 pc per side, with a resolution of 0.047 pc. In this domain, we impose an ISM with a constant number density of $n_0=0.2\, \mathrm{cm}^{-3}$, a temperature of 1000 K, and a magnetic field $B_0=2 \mathrm{ \mu G}$~\citep{jansson2012}, which are typical values of the Galactic warm neutral ISM \citep{Wolfire2008}.

As mentioned above, SNRs with jet-like morphologies can be sculpted by the interaction of a SN ejecta with an anisotropic CSM \ac{\citep{blondin1996,2020Galax...8...38C, alexandros21, ustamujic2021}}. In this work, we explore whether this type of scenario is also able to explain the origin of retangular remnants  by studying the effects of specific characteristics of the wind morphology. In this scenario our simulations are split into two steps. 
First, we model the formation and evolution of the stellar bubble and then, in the second simulation step, the evolution of the subsequent SNR within it.

Anisotropies in a CSM can be produced either by the stellar rotation and/or by the orbital motion of a close binary system \citep[e.g.][]{Bjorkman1993, Mastrodemos1999, Heger2000, Politano2011, gloria2020}. The angular momentum transportation from the parent stellar system to its mass outflows results in the formation of circumstellar structures characterised by a bipolar shape with a density enhancement at the equatorial plane. The effect of this mechanism is more dramatic for cases where the stellar wind velocity is comparable to the stellar/orbital rotation velocity.  Thus, bipolar circumstellar structures are frequently found surrounding stars that are characterised by dense and slow stellar winds such as Red Supergiants \citep[RSGs; ][]{asida1995, Tiffany2010} and stars at the Asymptotic Giant Branch \citep[AGB;][]{decin2020}. The present study will focus on these two types of stars.

To describe the angular  anisotropy of an equatorially confined stellar wind, we follow the equations of \citet{mellema1991} that provide the density distribution as a function of radius $r$ and polar angle $\theta$ in spherical coordinates: 
\begin{equation}
\rho_w(\theta, r)=\frac{\dot{M}_{\rm p}}{4\pi v_{\rm p} r^2}f(\theta),
    \label{eq:rhow}
\end{equation}
where $\dot{M}_{\rm p}$ and $v_{\rm p}$ are the mass-loss rate and the terminal velocity of the stellar wind at the poles ($\theta= 0$). The function $f(\theta)$ is given by:
\begin{equation}
    f(\theta)=\frac{1}{1-\alpha}\left[1-\alpha\frac{1-\exp{(-2\beta \cos^2\theta)}}{1-\exp{(-2\beta)}}\right],
    \label{eq:ftheta}
\end{equation}
where $0 \le \alpha < 1$ determines the equator-to-pole density ratio which is given by $(1-\alpha)^{-1}$, and $\beta$ determines the width of the equatorial region ($\theta\sim \pi/2$), i.e., $\beta<1$ ($\beta>1$) produces a wide (narrow) equatorial region. 

The wind velocity distribution also depends on the polar angle as:
\begin{equation}
    v_{\rm w}(\theta)=\frac{v_{\rm p}}{f^{1/2}(\theta)}.
    \label{eq:vel}
\end{equation}
In this way, the stellar wind ram pressure $\rho_{\rm w} v_{\rm w}^2$ is isotropic.

Based on previous works \citep[see e.g.][]{raga2008,Meyer2021}, we choose $\alpha=0.95$, which implies an equator-to-pole density ratio of 20.
The parameter $\beta$ is set as 0.1 or 5. In Figure \ref{fig:rbw_rhow}, we plot $\rho_{\rm w}(\theta)/\rho_{\rm {w,max}}$ for the chosen values of parameters $\alpha$ and $\beta$.
We can get the radius $R_{\rm b}(\theta)$ of the outer boundary of the stellar bubble by combining our Eqs. (\ref{eq:rhow})-(\ref{eq:vel}) into:
\begin{equation}
    \frac{R_{\rm b}(\theta_1)}{R_{\rm b}(\theta_2)}=\bigg[ \frac{\dot{M}(\theta_1)}{\dot{M}(\theta_2)}\bigg]^{1/5} \bigg[\frac{v_{\rm w}(\theta_1)}{v_{\rm w}(\theta_2)}\bigg]^{2/5},
\end{equation}
which is the Eq. (3) of \citet{alexandros21}, and replacing $\theta_1$ by $\theta$ and $\theta_2=0$ in the last equation. Finally, $R_{\rm b}(\theta)$ satisfies:
\begin{equation}
    \frac{R_{\rm b}(\theta)}{R_\textrm{b,max}}=[f(\theta)]^{-1/10}
    \label{eq:rb},
\end{equation}
where $R_\textrm{b,max}=R_b(\theta=0)$ is the radius of the  wind bubble at the pole. Figure \ref{fig:rbw_rhow} displays $R_{\rm b}(\theta)/R_\textrm{b,max}$ for $\beta=5$ (blue line) and $\beta=0.1$ (orange line).  The plot corresponding to $\beta=5$ shows a peanut-like shape while the other case gives an oval morphology.

Integrating the mass flow rate $\rho_{\rm w} v_{\rm w}$ over an sphere of radius $r$, we obtain the total mass loss rate $\dot{M}_{\textrm{tot}}$ from Eqs.(\ref{eq:rhow})--(\ref{eq:vel}) as:
\begin{eqnarray}
  \dot{M}_{\text{tot}} & = & \int_0^{2 \pi} \int_0^{\pi} \rho_{\rm w} (\theta) v_{\rm w}
  (\theta) r^2 \sin \theta\, d \theta\, d \phi = \nonumber\\
  & = & \int_0^{2 \pi} \int_0^{\pi} \frac{\dot{M}_{\rm p} f
  (\theta) }{4 \pi v_{\rm p} r^2} \frac{v_{\rm p}}{f^{1/2} (\theta)} r^2 \sin \theta\, d \theta\, d \phi = \nonumber\\
  & = & \frac{\dot{M}_{\rm p}}{2} \int_0^{\pi} f^{1/2} (\theta) \sin \theta\, d
  \theta =\frac{\dot{M}_{\rm p}}{2}  I_{1/2} (\alpha , \beta)
  \label{eq:mdtot}
\end{eqnarray}
where $I_{1/2} (\alpha, \beta)$ is the result of the integral:
\begin{equation}
  I_{1/2}(\alpha,\beta) =\int_0^{\pi} \left\{ \frac{1}{1 - \alpha} \left[ 1 - \alpha \frac{1 -
  \exp (- 2 \beta \cos^2 \theta)}{1 - \exp (- 2 \beta)} \right] \right\}^{1/2} \sin \theta\, d \theta  \label{eq:itheta}
\end{equation}
and must be obtained numerically for each choice of $\alpha$ and $\beta$.
Results for a few choices for the isotropic case are shown in
Table \ref{tab:par} (4th column).
\begin{table}
	\centering
        \begin{tabular}{ccccc} 
		\hline
		Run & $\beta$$^{(a)}$ & $\dot{M}_{\text{tot}}$ (${\textrm M}_\odot \textrm {yr}^{-1}$)$^{(b)}$ &
		$I_{1/2}$($\alpha$,$\beta$)$^{(c)}$ & $M_\textrm{wb}(M_\odot)$$^{(d)}$\\
		\hline
		r1a & 5.0 & $5\times 10^{-6}$& 4.4 & 1.5\\
		r1b & 0.1 & $5\times 10^{-6}$ & 7.1 &1.5\\	
	    r2a & 5.0 & $1\times 10^{-5}$& 4.4 & 3.0\\
		r2b & 0.1 & $1\times 10^{-5}$& 7.1 & 3.0\\
		r3a & 5.0 & $2\times 10^{-5}$ &4.4 & 6.0\\
		r3b & 0.1 & $2\times 10^{-5}$ &7.1 & 6.0\\
		\hline
	\end{tabular}
\caption{Stellar wind parameters: $(a)$ parameter used in Eq.(\ref{eq:ftheta}); $(b)$ stellar wind total mass-loss rate; $(c)$ integral value given by Eq.(\ref{eq:itheta}); $(d)$ total mass ejected by the stellar wind.}
	\label{tab:par}
\end{table}

In light of this calculation, it is more convenient to define the wind density
distribution in terms of the total mass-loss rate, $\dot{M}_{\text{tot}}$,
instead of the value at the pole, $\dot{M}_{\rm p}$ (see eq. (\ref{eq:rhow})):
\begin{equation}
  \rho_{\rm w} (\theta, r) =\frac{\dot{M}_{\text{tot}}}{2 \pi I_{1/2} (\alpha,
  \beta) v_{\rm p} r^2} f (\theta)
  \label{eq:rhow2}
\end{equation}
where $I_{1/2} (\alpha, \beta)$ is given by Eq.
(\ref{eq:itheta}).


During the first phase of the simulations, we impose the stellar wind condition considering an inner boundary, given by a spherical surface with radius $R_{\rm w}=0.28$~pc (centred in the middle of the computational domain), where the wind's density and velocity have a constant income into the surrounding medium according to the mass-loss rate chosen. We have set the parameter $\beta$ as 0.1 or 5. To avoid possible numerical artefacts stemming from the Cartesian grid, we tilt the polar axis of the stellar wind (which is contained in the $x=0$ plane) by 60$^{\circ}$ concerning the $z$-axis so that the polar direction of the wind does not coincide with any of the simulation axes. 
Besides, the ambient magnetic field is on the $xy$ plane, making an angle of 30$^{\circ}$ with the $y$-axis. 

We performed six runs imposing stellar winds with a $\dot{M}_\textrm{tot}$ of 0.5, 1, or $2\times 10^{-5} ~ \textrm{M}_{\odot}\, \textrm{yr}^{-1}$ (see Table \ref{tab:par}). In all models the polar wind terminal velocity is $v_{\rm p}=25~ \mathrm{km\, s^{-1}}$ and $\alpha=0.95$ (as mentioned above). The adopted wind mass  loss rates and terminal velocities are typical for RSGs \citep[e.g.][]{Beasor2020},  and AGB stars \citep[e.g.][]{Vassiliadis1993} studied in this work. We let all runs evolve for 300~kyr to form the wind bubble, a value representative of the duration of these stellar winds \citep{hofner2018,Hernandez2019}. During this elapsed time with the given $\dot{M}_\textrm {tot}$, the stellar winds inject a total mass $M_\textrm{wb}$ from 1.5 to 6 $\mathrm{M}_\odot$ into their surrounding environment. In table \ref{tab:par}, we list the parameters employed in all runs.

Once the stellar wind bubble forms, we impose conditions consistent with type Ia or II SN explosions in a sphere of radius $R_0=0.28$~pc at the centre of the computational domain. For both kinds of SNe, we choose an initial energy $E_0$ of $10^{51}$~erg \citep{martinez2022}. A fraction $f_{\rm K}=0.95$ of $E_0$ corresponds to the kinetic energy. The mass $M_{*}$ ejected by the Type Ia SN was chosen as 1.38$\mathrm{M}_\odot$, while this initial mass was 10$\mathrm{M}_\odot$ for the Type II SN. With these parameters, we can estimate an SNR initial age of $\simeq 10$~yr \citep{truelove1999}. Furthermore, we consider the CSM swept mass $M_\textrm{sw}$ by the initial SNR, integrating the Eq.(\ref{eq:rhow2}) in a sphere of radius $R_0$. Table \ref{tab:SNmass} summarises the resulting $M_\textrm{sw}$.
\begin{table}
    \centering
    \begin{tabular}{cc}
    \hline
    Run & $M_\textrm{sw} (M_\odot )$ \\
    \hline     
    r1a     &  0.15\\
    r1b     &  0.18\\    
    r2a     &  0.30\\
    r2b     &  0.35\\
    r3a     &  0.60\\    
    r3b     &  0.70\\
    \hline     
    \end{tabular}
            \caption{CSM swept up mass by the initial SNR}
    \label{tab:SNmass}
\end{table}
Then, the initial remnant has a  mass $M_{0}=M_*+M_\textrm{sw}$ and a constant density $\rho_{\rm c}$ up to a radius $r_{\rm c}$. For $r_{\rm c}\leq r\leq R_0$, the density is $\rho_{\rm c} (r_{\rm c}/r)^7$ \citep{Jun1996}. This outer region (i.e. with $r\geq r_{\rm c}$) contains a fraction $X_{\rm m}$ of $M_0$. The remaining $(1-X_{\rm m})M_0$ is uniformly distributed in the sphere with radius $r_{\rm c}$. Besides, the radial velocity $v_r$ grows linearly with $r$ as $v_r=v_0 (r/R_0)$, being $v_0$ the velocity at $r=R_0$. We can determine the values of $r_{\rm c}$, $\rho_{\rm c}$, and $v_0$ after integrating the remnant density and the kinetic energy density into a sphere of radius $R_0$. We obtain:
\begin{eqnarray}
  r_c&=&R_0 \bigg[\frac{1-(7/3)X_m}{1-X_m}\bigg]^{1/4}, \\
  \rho_c&=&\frac{3 M_0}{4\pi R^3_0}\frac{(1-X_m)^{7/4}}{(1-(7/3) X_m)^{3/4}},\ \text{and}\\
  v_0 &=& \sqrt{\frac{4 f_k E_0}{3 M_0 (1-X_m) y^2_r(7/5-y^2_r)}},
\end{eqnarray}
where in the last equation, $y_r=r_{\rm c}/R_0$. In our runs, we set $X_m=0.4$.\footnote{There is no difference if we start the remnant with the density profile given by the equations above or if we use a constant density profile.}


\subsection{The code}

\begin{figure*}
	\includegraphics[width=14cm]{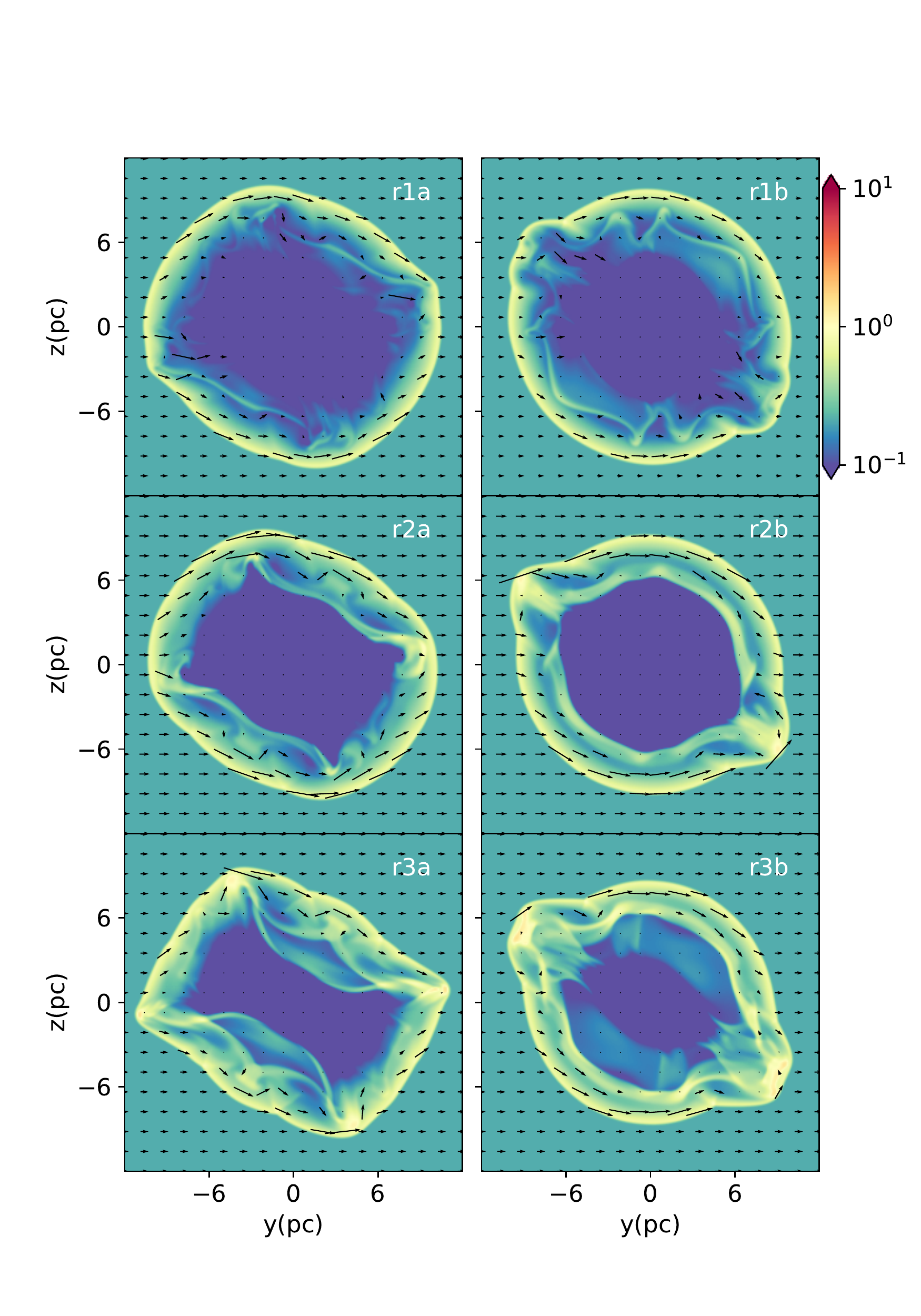}
    \caption{Density distributions maps ($x=0$ plane) of the SNR evolution for runs with $\beta=5$ (left column) and $\beta=0.1$ (right column). All maps correspond to an integration time of 2.5 kyr after SN explosion. In this case, we considered a Type Ia SN explosion. Black arrows display the magnetic field distribution. The logarithmic colour scale gives the density in $2\times 10^{-24}$ g~cm$^{-3}$ units. 
    Both axes units are given in pc.}
    \label{fig:comp_dens}
\end{figure*}

\begin{figure*}
	\includegraphics[width=14cm]{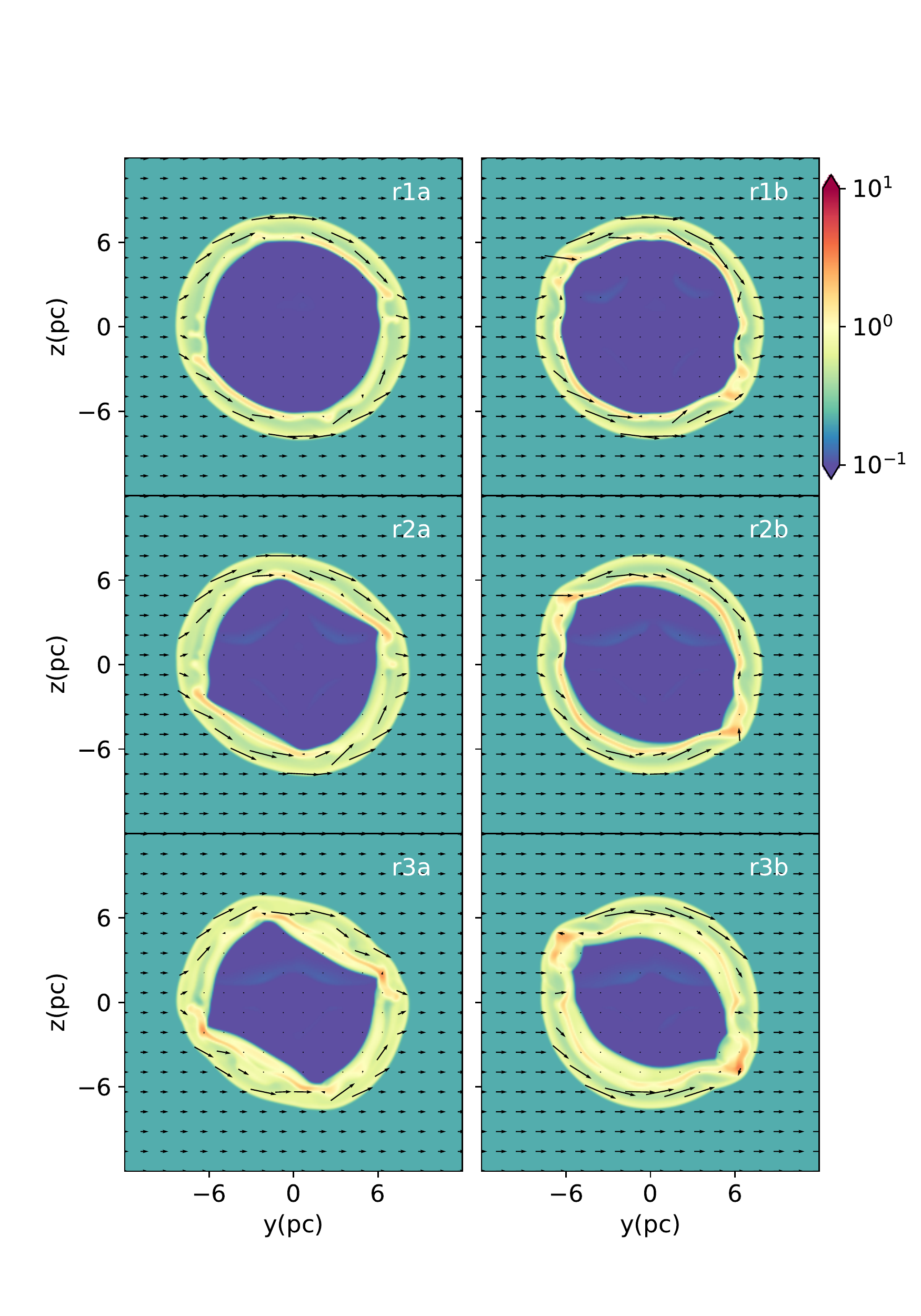}
    \caption{The same as Fig. \ref{fig:comp_dens} but for the case of a Type II SN event. The label on each map indicates the corresponding run. The logarithmic colour scale gives the density in g~cm$^{-3}$ units. Both axes units are given in pc.}
    \label{fig:comp_densII}
\end{figure*}
The numerical study was performed with the parallel 3D magneto-hydrodynamical (MHD) code {\sc guacho} \citep{Esquivel2009,Villarreal2018}. This code solves the ideal MHD equations in a fixed Cartesian grid :

\begin{equation}
    \frac{\partial\rho}{\partial t}+\nabla\cdot(\rho\mathbfit{u})=0\;,
	\label{eq:mass}
\end{equation}

\begin{equation}
    \frac{\partial(\rho\mathbfit{u})}{\partial t}+\nabla\cdot\left[\rho\mathbfit{u}\otimes\mathbfit{u}+\mathbfss{I}\left(p+\frac{B^2}{8\pi}\right)-\frac{\mathbfit{B}\otimes\mathbfit{B}}{4\pi}\right]=0\;,
	\label{eq:momentum}
\end{equation}

\begin{equation}
    \frac{\partial e}{\partial t}+\nabla\cdot \left[
    \left(e + p + \frac{B^2}{4\pi}\right)\mathbfit{u}-\left(\mathbfit{u} \cdot \mathbfit{B}\right) \mathbfit{B}
    \right]=Q_L\;,
	\label{eq:energy}
\end{equation}

\begin{equation}
    \frac{\partial \mathbfit{B}}{\partial t}-\nabla\times \left(\mathbfit{u}\times \mathbfit{B}\right)=0\;,
   	\label{eq:induction}
\end{equation}

\noindent where $\rho$, $\mathbfit{u}$, $p$, $\mathbfit{B}$ and $e$ are the mass density, velocity, gas pressure, magnetic field and total energy density, respectively. In Eq. (\ref{eq:momentum}), $\mathbfss{I}$ is the identity matrix. The energy density is given by $e=\rho u^2/2+p/(\gamma-1)+B^2/8\pi$, with $\gamma$ being the heat capacity ratio of the gas, which was set to 5/3. The code includes radiative cooling $Q_L$ (see Eq. \ref{eq:energy}) as a parameterised function of the temperature. We use the function for optically-thin cooling described in \citet{Dalgarno1972}. A second-order Godunov method with the approximate Riemann solver HLLD \citep{miyoshi2005}, was used to advance Eqs. (\ref{eq:mass})--(\ref{eq:induction}) in time. A zero-gradient (outflow) condition is imposed in all of the domain boundaries.

\subsection{Synthetic emission maps}
We consider two reference systems for performing synthetic emission maps from our numerical results. The first is associated with the computational domain ($xyz$ frame), while the second is given by the plane of the sky ($x^\prime y^\prime$ plane) and the line of sight (LoS, $z^\prime$ coordinate). At the beginning both frames coincide. We then rotate the computational domain relative to the $x^\prime y^\prime z^\prime$ frame to obtain maps with with different lines of sight.
\subsubsection{Synchrotron  emission}
\begin{figure*}
	\includegraphics[width=14cm]{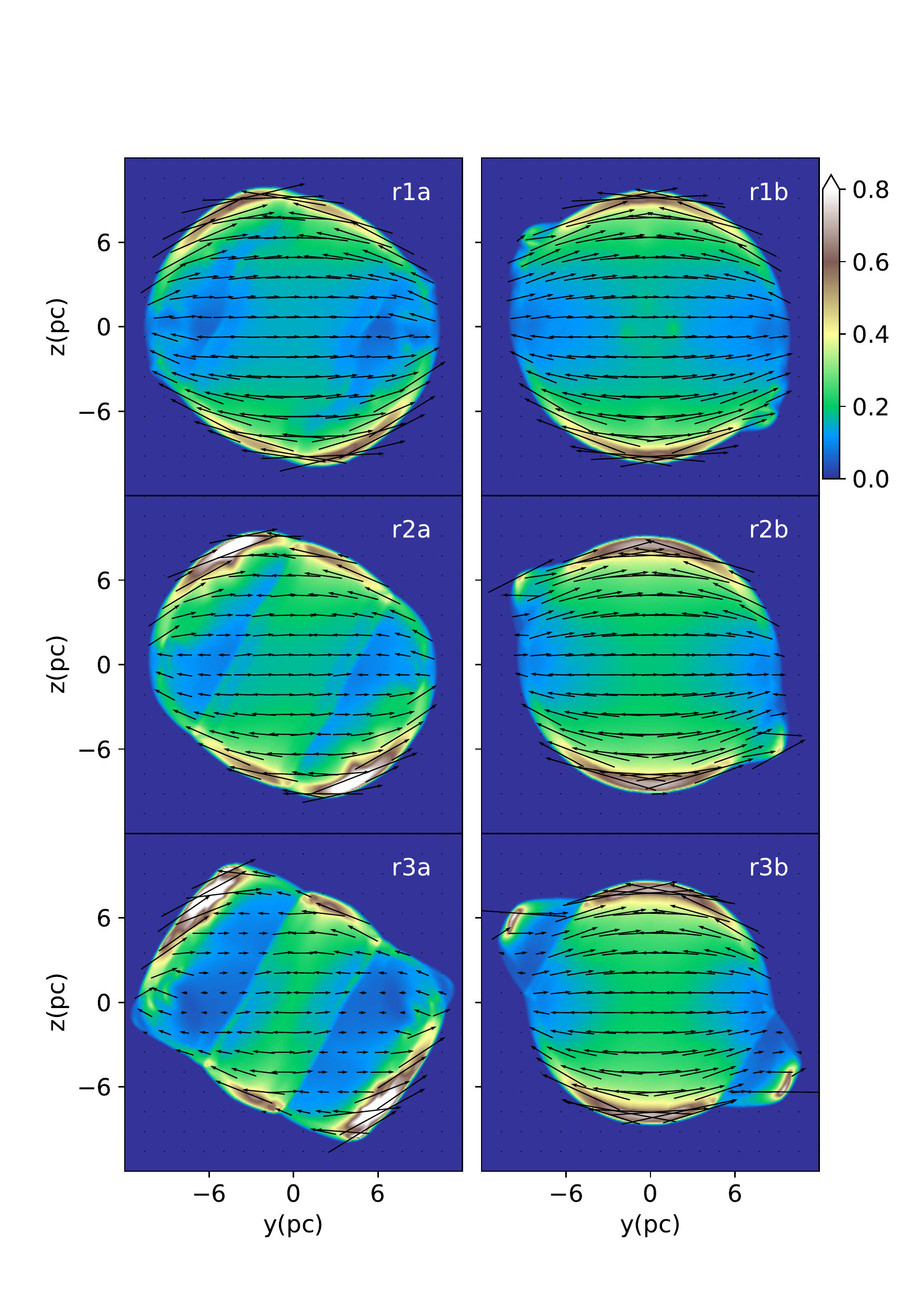}
    \caption{Synchrotron emission maps at 2.5~kyr after a Type Ia SN explosion for all runs.  The colour scale represents the synchrotron emission in arbitrary units overlaid with the magnetic field orientation, given by the Stokes parameters (black arrows). The LoS is the $\hat{x}$-axis. The horizontal and vertical axes units are in pc.}
    \label{fig:comp_sincro}
\end{figure*}

\begin{figure*}
	\includegraphics[width=14cm]{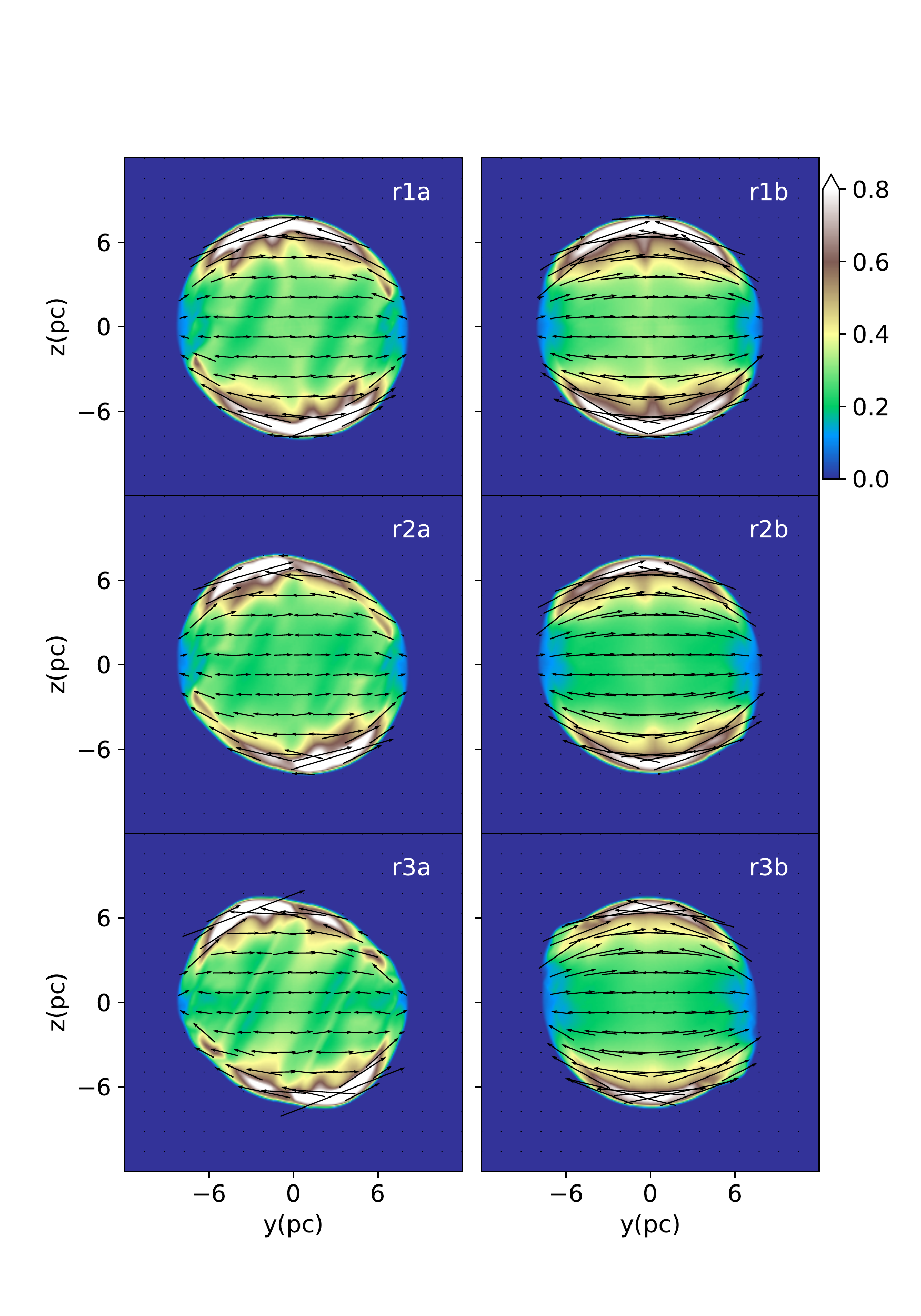}
    \caption{The same as Fig. \ref{fig:comp_sincro} but considering a Type II SN explosion. The LoS is the $\hat{x}$-axis. The horizontal and vertical axes units are in pc.}
    \label{fig:comp_sincroII}
\end{figure*}

After rotating the reference system, we calculate the synchrotron emissivity for each cell of our computational grid using the Stokes parameters $I$, $Q$, and $U$ for this purpose. The main aspects of the calculation are summarised below. Further details can be found in \citet{Aroche2020} and \citet{Meyer2021}.

The synchrotron specific intensity for each point $(x^\prime,y^\prime,z^\prime)$ in the rotated computational domain can be written as \citep{Jun1996}:
\begin{equation}
j_{\rm s}(x^\prime,y^\prime,z^\prime,\nu)=\kappa p^{2s}\rho^{1-2s} B^{s+1}_{\perp}\nu^{-s}, 
\label{eq:js}
\end{equation}  
where $p$ and $\rho$ are the pressure and density of the gas, $\nu$ is the observed frequency, $B_{\perp}$ is the magnetic field component perpendicular to the line of sight (LoS), and $s$ is the spectral index. The parameter $\kappa$ is a constant since we consider an isotropic particle acceleration mechanism.

To obtain synthetic synchrotron emission maps, we compute the total intensity of the synchrotron emission
as (see \citealt{Aroche2020}): 
\begin{equation}
I(x^\prime,y^\prime,\nu)=\int_\textrm{LoS}j_{\rm s}(x^\prime,y^\prime,z^\prime,\nu)dz^\prime, 
\label{eq:I}
\end{equation}  

We compute the Stokes parameters $Q$ and $U$ from the specific intensity as:
\begin{equation}
Q(x^\prime,y^\prime,\nu)=\int_\textrm{LoS}f_{\rm p}\,j_{\rm s}(x^\prime,y^\prime,z^\prime,\nu)\cos(2\phi)dz^\prime,
\label{eq:Q}
\end{equation}  
\begin{equation}
U(x^\prime,y^\prime,\nu)=\int_\textrm{LoS}f_{\rm p}\,j_{\rm s}(x^\prime,y^\prime,z^\prime,\nu)\sin(2\phi)dz^\prime,
\label{eq:U}
\end{equation}
with $\phi$ the position angle of the local magnetic field in the plane of the sky, and  $f_{\rm p}$ is the linear polarization degree, related to the spectral index $s$, viz:
\begin{equation}
f_{\rm p}=\frac{s+1}{s+5/3}.
\end{equation}  

Finally, the magnetic field position angle ($\Phi_B$) is obtained by, 
\begin{equation}
    \Phi_B(x^\prime,y^\prime,\nu)=\frac{1}{2}\arctan\bigg( \frac{U(x^\prime,y^\prime,\nu)}{Q(x^\prime,y^\prime,\nu)}\bigg)
\end{equation}

\subsubsection{Thermal X-ray emission}

We calculate the X-ray emission coefficient for each cell of the rotated system, assuming a low-density regime, as:
\begin{equation}
  j_\mathrm{x}(x^\prime,y^\prime,z^\prime)=n_{\rm e}^2\xi(T), \label{eq:jrx} 
\end{equation} 
where $n_{\rm e}$ is the electron density (equal to $n$, the gas density), $T$ the temperature computed from our numerical simulations (assuming that the electronic and ionic temperatures are the same), and $\xi(T)$ is a smooth function of temperature. The $\xi(T)$ function was computed for the range [0.1-10] keV and a solar metallicity using the CHIANTI atomic database (\citealt{Dere1997,Landi2006A}) and employing the ionization equilibrium given by \citet{Mazzotta1998}. Furthermore, we take into account the interstellar absorption considering a column density of $N_{\rm H}=5\times10^{21}$ cm$^{-2}$.
To obtain the synthetic map, we integrate $j_\mathrm{x}(x^\prime,y^\prime,z^\prime)$ along $z^\prime$ (i.e., the LoS):
\begin{equation}
    I_{\mathrm{x}}(x^\prime,y^\prime,z^\prime)=\int_\mathrm{LoS} j_\mathrm{x}(x^\prime,y^\prime,z^\prime) dz^{\prime}
    \label{eq:irx}
\end{equation}
\section{Results}\label{Sec:3}
\begin{figure*}
	\includegraphics[width=14cm]{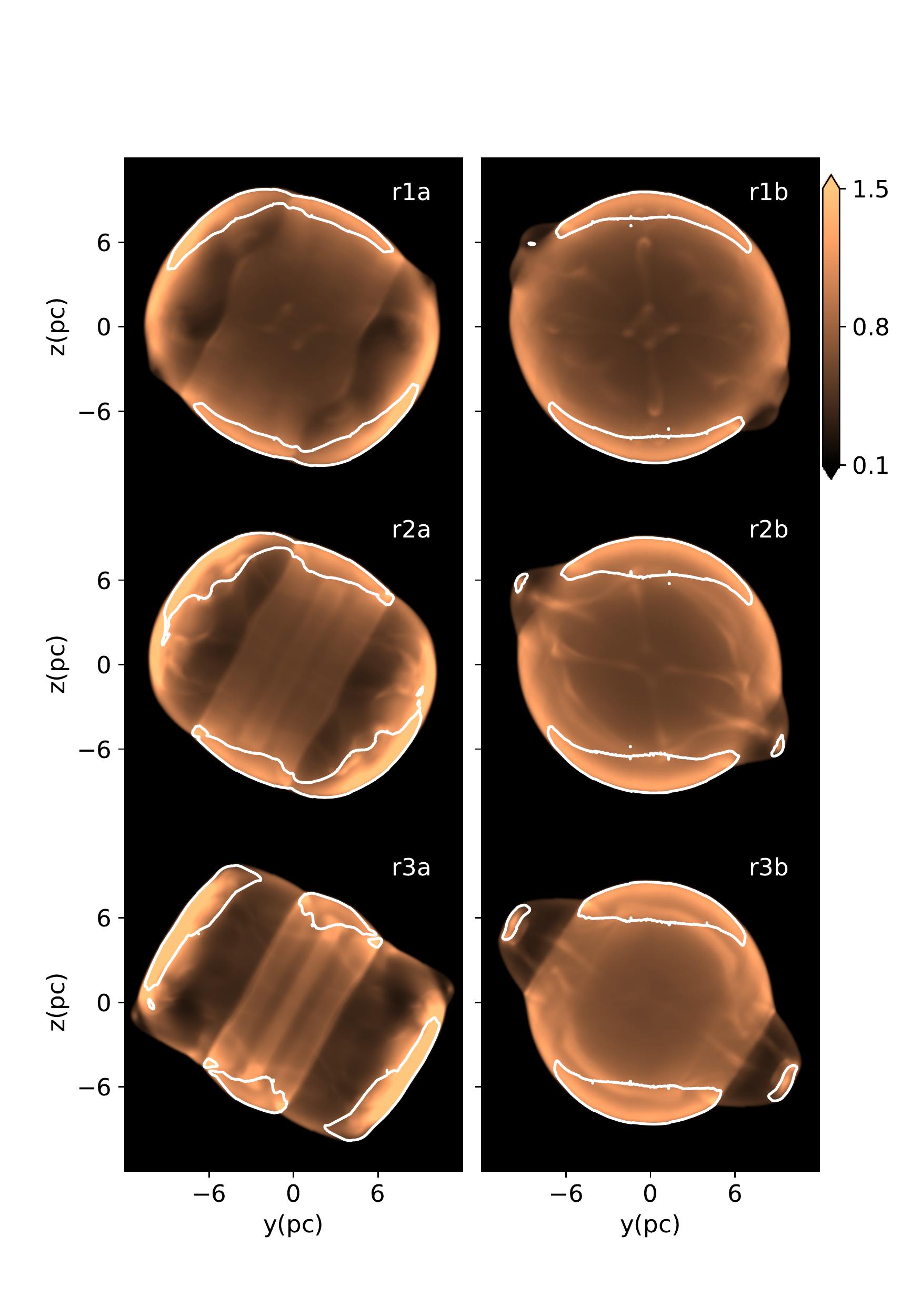}
    \caption{Thermal X-ray emission maps, which are displayed in units of $10^{-6} \mathrm{erg\, s^{-1}\, cm^{-2}\, ster^{-1}}$, obtained for all runs. The integration time is 2.5~kyr after a Type Ia explosion. White contours enclose regions with strong synchrotron emission. The $\hat{x}$ axis is the line of the sight.
    The horizontal and vertical axes units are in pc.
    }
    \label{fig:comp_xray}
\end{figure*}

\begin{figure*}
	\includegraphics[width=14cm]{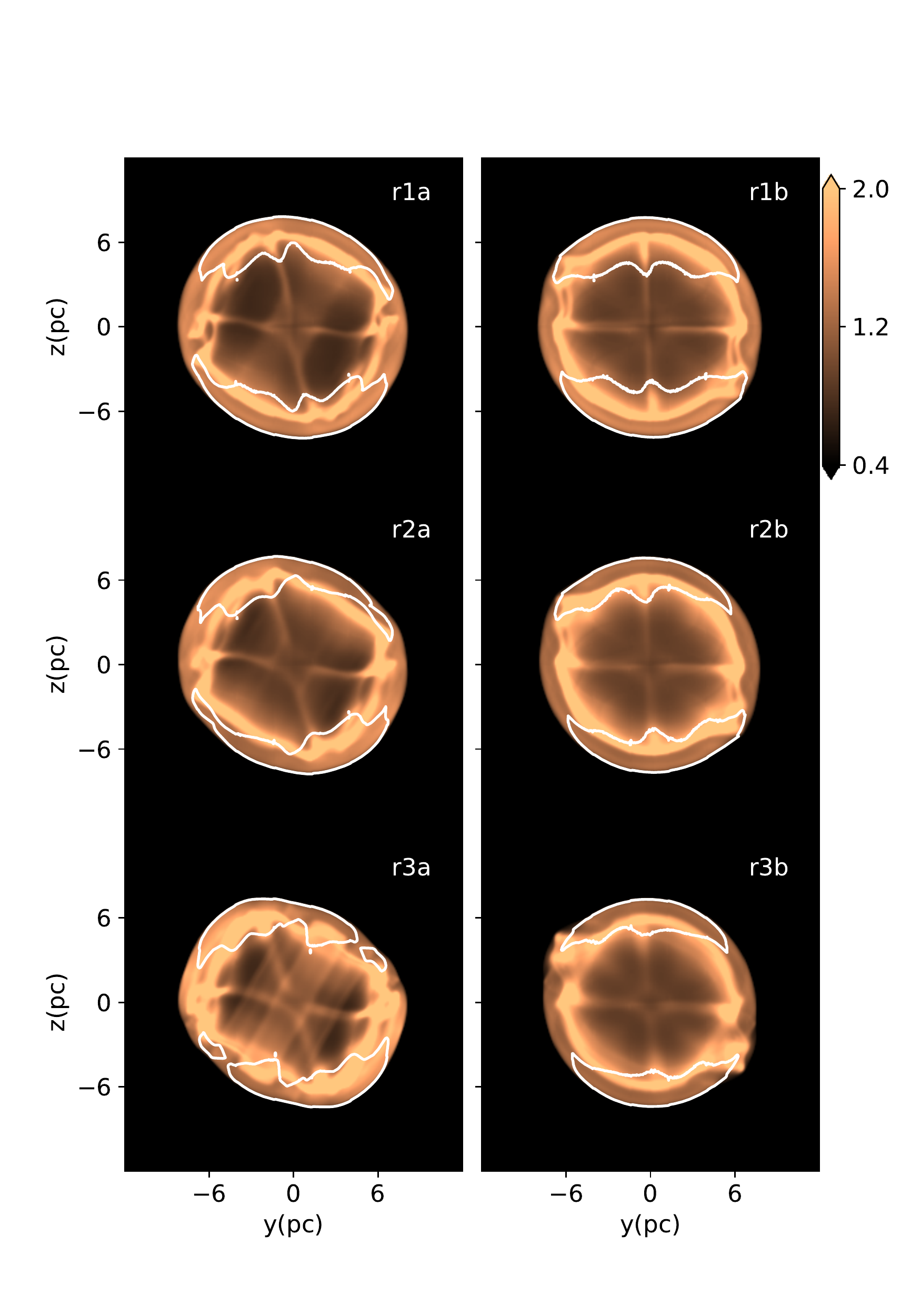}
    \caption{Same as Figure \ref{fig:comp_xray} but for the case of a Type II SN event.  White contours enclose bright synchrotron regions. The $\hat{x}$ axis is the LoS.
    The horizontal and vertical axes units are in pc.
    }
    \label{fig:comp_xrayII}
\end{figure*}
\begin{figure*}
    \centering
    \includegraphics[width=14cm]{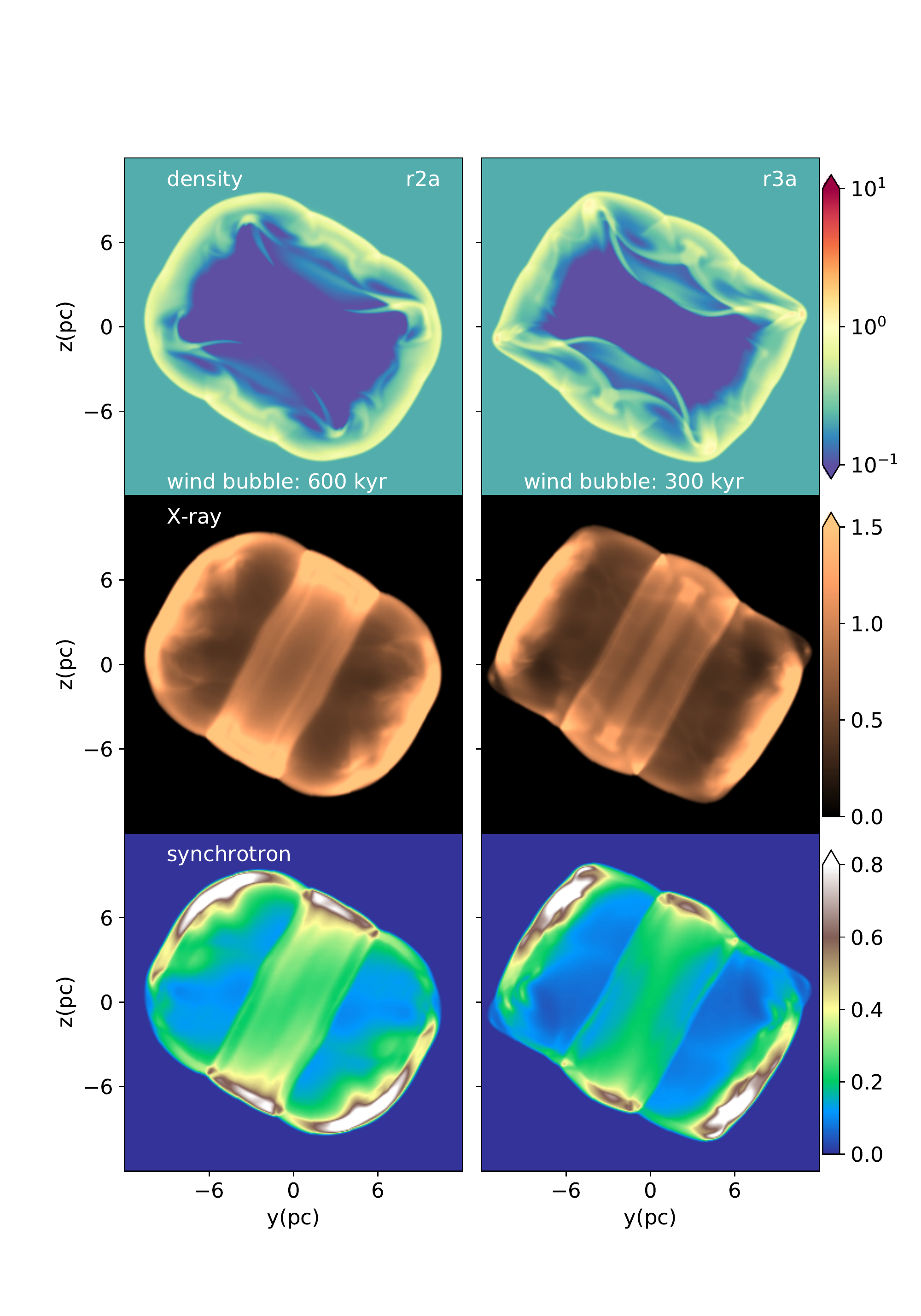}
    \caption{Comparison of the density distribution on the $X=0$ plane (in $2\times 10^{-24}\, \mathrm{g\, cm^{-3}}$ units, top row), X-ray (given in $10^{-6}\mathrm{erg\, cm^{-2}\, s^{-1}\, ster^{-1}}$, middle row), and synchrotron (in arbitrary units, bottom row) maps obtained for runs r2a (left column, but with a stellar wind bubble age of 600 kyr) and r3a (right column). All these maps correspond to a type Ia explosion and the progenitor injects 6 M$_\odot$ into its surrounding medium. The integration time is 2.5~kyr.}
    \label{fig:comp_rect}
\end{figure*}

\begin{figure*}
    \centering
    \includegraphics[width=16cm]{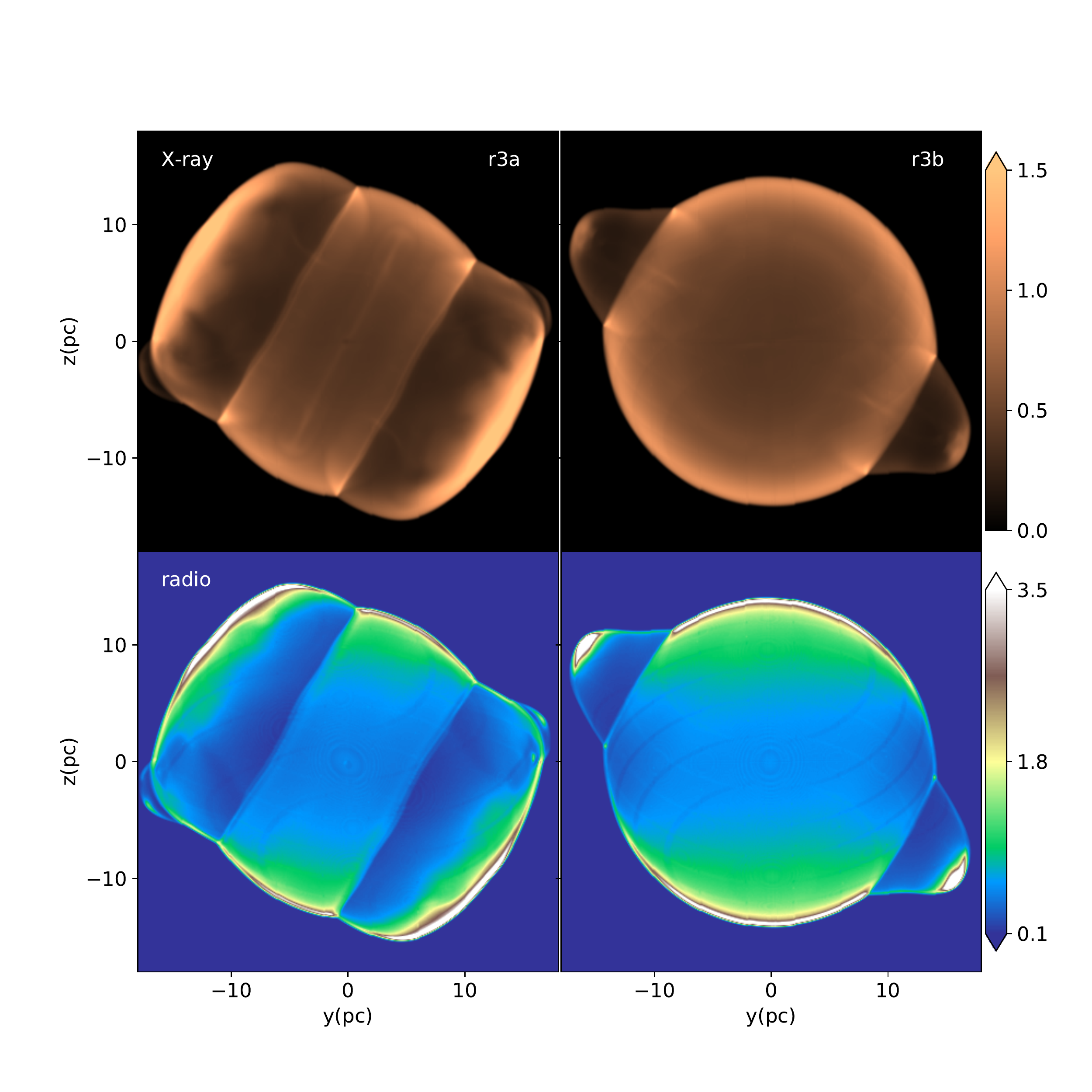}
    \caption{Comparison between X-ray emission (upper row) and synchrotron emission maps (bottom row) obtained for run r3a (left column) and r3b (right column). All maps correspond to an integration time of 6.5 kyr, after SN explosion.  The X-ray emission is given in units of $5\times10^{-6}\mathrm{erg\, s^{-1}\, cm^{-2}\, ster^{-1}}$ and synchrotron emission in arbitrary units. Maps in each row use the same colour scale.}
    \label{fig:comp_obs}
\end{figure*}

\begin{figure*}
    \centering
    \includegraphics[width=16cm]{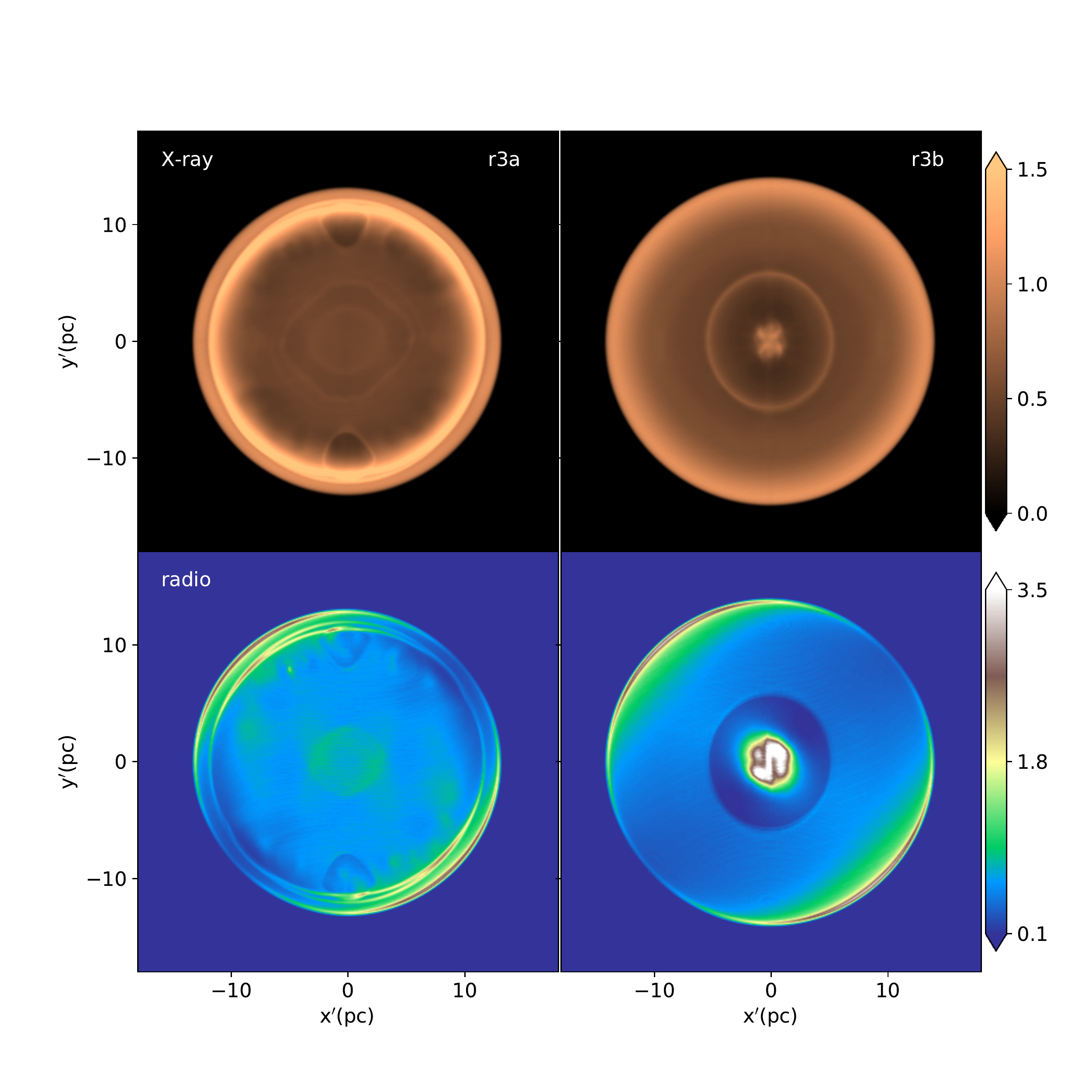}
    \caption{Same as Fig. \ref{fig:comp_obs} but considering that the LoS is the polar wind axis.}
    \label{fig:comp_obs_pol}
\end{figure*}

\subsection{Density distribution}

Figure \ref{fig:winds} illustrates the density distribution (colour maps) and the magnetic field (black arrows) on the $x=0$-plane for the stellar wind bubbles of all runs (indicated by the label on the top-right corner of each panel), at an integration time of $300~\textrm{kyr}$. Left (right) column maps correspond to a stellar wind with a narrow, dense equatorial region (wide, dense equatorial region). The total mass-loss rate of the stellar wind increases from the top row to the bottom, with the bubble size growing accordingly. The wind bubbles obtained for runs r1a, r2a, and r3a (left column, Figure \ref{fig:winds}) exhibit an hourglass shape, which differs from the peanut-like morphology obtained from Equation (\ref{eq:rb}); see the blue line in Figure \ref{fig:rbw_rhow}. This difference is due to the interstellar magnetic field, which allows gas motion along its direction and restrains it along its normal direction \citep{meyer_mnras_464_2017}. We can verify this by calculating the plasma parameter, defined as the thermal and magnetic pressure ratio ($\beta_p=(8\pi p)/B^2$). In the case of stellar bubbles, the value of this parameter is of the order of $10^{-3}$, which shows that the magnetic field has a dominant role in the fluid dynamics. Instead, ovoid-shaped stellar wind bubbles appear for runs r1b, r2b, and r3b (right column maps of Figure \ref{fig:winds}). For these last cases, the dense, broad equatorial region favours the gas motion towards the stellar wind poles.

For all models, at this stage we impose a type Ia and a type II SN explosion at the centre of the grid. Then, we let the SNR evolve and interact with the previously formed circumstellar bubble. Figures \ref{fig:comp_dens} and \ref{fig:comp_densII} show the density distribution maps obtained for type Ia and II SN explosions, respectively. All these maps correspond to an integration time of 2.5 kyr after the SN event.

The left column of Figure \ref{fig:comp_dens} displays those models with $\beta=5$, showing an elongated shell morphology (upper and middle maps of this column) or a rectangular shape (bottom map). In addition, small ear-like features appear in the equatorial region (see the bottom-left map of Figure \ref{fig:comp_dens}), which agrees with the results of \citet{alexandros21}. This ear-like structure is a 2D projection of a dense equatorial belt.

The right-column maps of Figure \ref{fig:comp_dens} show the density maps obtained for the $\beta=0.1$ models. Run r1b produces a spherical shell-like remnant with tiny protrusions in the polar regions. Instead, runs r2b and r3b display almost spherical remnant shells with well-developed ear- or jet-like features at both poles. Similar results were obtained by \citet{blondin1996} and \citet{ustamujic2021}.

Figure \ref{fig:comp_densII} shows the maps generated for the case of a type II supernova explosion.  As expected, the resulting SNRs reveal -qualitatively- similar morphologies to those obtained by the type Ia explosion since the same physical processes sculpted the final SNR morphology. Nevertheless, given than in this case the SN ejecta mass is larger that the total circumstellar mass, the effect of the SN-CSM interaction on the resulting SNR is  mitigated.  In particular, maps of runs r1a, r1b, r2a, and r2b show nearly spherical remnants, while the density map of run r3a exhibits a rectangular shape with rounded edges and elongated in the polar axis. On the other hand, we observe a spherical remnant with small opposing ears in the polar regions for run r3b.

\subsection{Synchrotron and thermal X-ray emission}

In order to make a more concrete comparison between the simulations and observations beyond the density distribution maps, we computed synthetic non-thermal radio and thermal X-ray emission maps from our numerical results.

Figures 6 and 7 show the synchrotron emission maps performed for all runs, considering type Ia and II supernova explosions, respectively. To obtain these maps, we made two rotations, both by an angle of -90 degrees: the first one around the x-direction, and the second, around the y-direction. In this way, the plane of the sky is the yz-plane,  and the LoS is the x-direction.

The top-left panel of Figure 6 displays a bilateral morphology for the supernova remnant for run r1a, where the SNR shell appears slightly elongated in the polar wind direction. The synthetic remnant of run r2a  (middle left map of Figure 6) 
shows an emitting band crossing the centre of the SNR. A higher $\dot{M}_{tot}$, run r3a,  produces the rectangular shape observed on the density maps (Fig. \ref{fig:comp_dens}) with a three-band configuration in the synchrotron emission (see bottom left panel of Figure 6).

The synchrotron emission, corresponding to the runs r1b, r2b, and r3b (right column of Figure 6), displays a spherical shape with a bilateral brightness distribution and protrusions along the polar direction. These protrusions grow in size and intensity with the increasing $\dot{M}_{\mathrm{tot}}$.

The synthetic synchrotron maps displayed in Figure 7 correspond to the case of a type II supernova. All maps show a bilateral morphology

Thermal X-ray emission maps obtained for all runs are shown in figures 8 and 9 for type Ia and II explosions, respectively. The maps in the top-row of figure 8 and all maps in figure 9 exhibit a shell-like shape. Instead, the maps of runs r2a and r3a in figure 8 show that the X-ray emission is distributed in three almost parallel stripes, while the maps of runs 2b and r3b (figure 8) display a shell-type morphology with two weak polar ear-like features.

\subsection{Analysis of the results}
The previous results show that the parameter that most affects the final  morphology of a SN is the stellar wind density distribution, which increases the density towards the wind equator. The parameter $\beta$ determines the shape of the equatorial region. For example, stellar wind bubbles with $\beta=0.1$ have dense, wide torus-producing spherical remnants with polar protrusions, such as those observed in the middle-right and bottom-right maps in Figure \ref{fig:comp_dens}, \ref{fig:comp_sincro}, and \ref{fig:comp_xray} resembling the appearance of SNR G290.1-0.8. Instead, the SNR morphology elongates in the polar direction for $\beta=5$, which generates a narrow equatorial torus. In this case, the SNR shape evolves from an elongated shell (top and middle maps, left column, Figures \ref{fig:comp_dens}, \ref{fig:comp_sincro}, and \ref{fig:comp_xray}) to a rectangular morphology (bottom-left map). The rectangular shape obtained in the last map resembles the appearance of some SNRs, such as G332.5-5.6. The mass-loss rate increases from top to bottom.

Another factor that modifies the SNR morphology is the $M_\textrm{wb}/M_0$ ratio. For example, the case for a type Ia SN explosion in the scenarios given by runs r2a, r2b, r3a, and r3b (see middle and bottom maps of figures \ref{fig:comp_dens}, \ref{fig:comp_sincro}, and \ref{fig:comp_xray}) correspond to $M_\textrm{wb}/M_0>1$. Instead, when $M_\textrm{wb}/M_0\simeq 1$ the remnant morphology is slightly affected, either in the case of a type Ia SN (runs r1a and r1b, see top maps of Figures \ref{fig:comp_dens}, \ref{fig:comp_sincro}, and \ref{fig:comp_xray}) or a type II event (bottom maps in Figures \ref{fig:comp_densII}, \ref{fig:comp_sincroII}, and \ref{fig:comp_xrayII}). The latter scenario involves a more massive circumstellar medium (CSM) and a Type II SN explosion. An RSG star can be the progenitor of a Type II event but does not generate a CSM that is massive enough for stars with initial masses $\leq 16\, \mathrm{M_{\odot}}$  \citep{beasor2018}. On the other hand, an AGB star does form a massive CSM \citep{villaver2007}, but it can not be the progenitor of a massive SN explosion. 
Thus, the high-mass CSM and a Type II SN scenario does not seem plausible. 
Finally, we do not observe an appreciable effect if $M_\textrm{wb}/M_0<1$ and a type II event occurrs (top and middle maps, Figures \ref{fig:comp_densII}, \ref{fig:comp_sincroII}, and \ref{fig:comp_xrayII}).

These findings lead us to wonder which is the determining factor in sculpting rectangular SNRs. To answer this question, we performed a new simulation with the initial setup of model r2a but letting the stellar bubble first evolve for 600 thousand years. This way, the stellar wind injects the same mass as run r3a. Inside this bubble, we imposed a Type Ia SN. Figure \ref{fig:comp_rect} compares the morphology and characteristics of the supernova remnant obtained from this new simulation with the one from run r3a. This figure shows the density distribution, X-ray emission, and synchrotron emission maps for an integration time of 2.5 kyr, showing that both remnants are approximately the same size. In addition, we observe three bands on both the X-ray and synchrotron emissions. The morphology of the SNR for the new simulation is almost rectangular, albeit with slightly rounded corners. Both simulations produce synthetic maps which match well with SNR G332.5-5.6 observations in both frequency regimes. So, we can conclude that rectangular morphologies are produced when the two following conditions occur simultaneously:  (1) a dense and narrow equatorial region ($\beta=5$), and (2)  a massive CSM (i.e., $M_\textrm{wb}>M_0$). To form jet-like morphologies in SNRs, the first condition must be replaced by a dense, broad equatorial region ($\beta=0.1$).

\subsection{Comparison with observed SNRs}

As a final test, we investigate if the rectangular or jet-like morphologies survive after long times. We carried out two new simulations employing the initial setups of runs r3a and r3b but with a larger computational domain size (36 pc)\footnote{Also, the initial stellar wind and SNR radii were increased to 0.42 pc to keep the same sizes in pixels as in the simulation with a computational domain with 24 pc per side}. The SNRs evolve by 6.5 kyr, and the resulting synthetic emission maps are displayed in Figure \ref{fig:comp_obs}. The X-ray and synchrotron maps of run r3a show a three-band configuration with an external rectangular shape. Along the polar direction, the remnant has a size of 30 pc, coincidentally with the size of G332.5-5.6 \citep{reynoso2007}. Besides, Figure 17 of \citet{stupar2007} shows that the radio and X-ray emissions in this SNR are both roughly distributed throughout three parallel bands, in reasonable agreement with our results.

On the other hand, the new simulation for run r3b (right-bottom map of Figure \ref{fig:comp_obs}) agrees with the jet-like radio morphology observed for the SNR G290.1-0.8 \citep{reynoso2006,milne1989}. Furthermore, this simulation produces a synthetic remnant with the observed physical size of 38 pc \citep{reynoso2006}, which coincides with its actual physical size along the polar direction. Furthermore, the simulated thermal X-ray emission in the right-top map of Figure \ref{fig:comp_xray}  displays morphology and brightness distributions, which are in qualitative agreement with the observations of \citet{Seward1990} and \citet{garcia2012} for this object.

Finally, we should point out that the final SNR morphology depends on the observer's point of view. To illustrate this, Figure \ref{fig:comp_obs_pol} shows the same cases as in Figure \ref{fig:comp_obs}, considering the polar axis as the LoS. In Figure \ref{fig:comp_obs_pol}, thermal X-ray maps (top row) show shell-like SNRs, while synchrotron maps (bottom row) reveal barrel-like morphologies. Some central emission also appears in the maps for run r3b, corresponding to the ear-like features now observed along the LoS.

\section{Conclusions}\label{Sec:4}

In this work, we present 3D MHD simulations of the evolution and emission of an SNR expanding within the stellar wind bubble driven by its progenitor star (with the main sequence mass $\leq 16~ \mathrm{M}_\odot$). We consider an anisotropic stellar wind, with a higher density towards the equatorial region than towards the poles. We aim to examine the conditions that produce jet-/ear-like and rectangular morphologies considering this framework.

We ran several simulations varying the stellar wind mass-loss rate, the $\beta$ parameter (which controls the shape of the equatorial region of the wind), and the SNR explosion type.

We find that the bubble formed by the wind from the progenitor star has little effect on an SNR if its total mass $M_\textrm{wb}$ is less than the mass ejected by the supernova, $M_0$. For example, this situation can occur for the wind of an RSG and a type II SN explosion (Figures \ref{fig:comp_densII}, \ref{fig:comp_sincroII}, and \ref{fig:comp_xrayII}). According to the work of \citet{beasor2018} and \citet{zapartas2021}, an RSG star can inject no more than about $3~ \mathrm{M}_\odot$ while a type II SN can release $10~\mathrm{M}_\odot$ into the
surrounding medium for a progenitor star with initial $\leq 16~\mathrm{M}_\odot$.

Instead, if the opposite case occurs (i.e., $M_\textrm{wb}>M_0$), the remnant evolution and its emission are strongly affected. This scenario is met, for example, when an AGB stellar wind is followed by a type Ia SN event. Besides, the $\beta$ parameter profoundly impacts the shape of the pre-SN wind bubble (Figure \ref{fig:winds}) and, consequently, the resulting SNR morphology (middle and bottom maps, left column, Figure \ref{fig:comp_dens}). Our results of runs r2a and r3a show that rectangular SNRs form when the equatorial region is a narrow, dense disk, as we observe in the bottom-left map of Figure \ref{fig:comp_dens} (see also Figure \ref{fig:comp_rect}). In contrast, SNRs with jet-like morphologies are found when the stellar wind is not as equatorially confined, corresponding to runs r2b and r3b (middle and bottom maps, right column in Figure \ref{fig:comp_dens}).

We performed synthetic synchrotron and thermal X-ray emission maps from our numerical results to compare with some SNRs belonging to these morphological classes. As a result, runs r2a and r3a, for the case of a type Ia SN, reproduce the global rectangular morphology and brightness distribution of the synchrotron and thermal X-ray emission observed (left maps in Figure \ref{fig:comp_obs}) for the SNR G332.5-5.6 \citep{reynoso2007,stupar2007}. For runs r2b and r3b, the simulations reproduce the main observed morphological features of the SNR G290.1-0.8 \citep{milne1989,Seward1990,reynoso2006}, which has a jet-like or ear-like shape (right maps in Figure \ref{fig:comp_obs}), and can also be helpful to represent other SNRs such as S147 \citep{ren2018} and G309.2-0.6 \citep{gaensler1998}.

In summary, 
we suggest that a scenario where a type Ia SN evolves within a cavity blown by an AGB anisotropic wind can produce rectangular morphologies, like for instance the SNR G332.5-5.6, provided the formation of a narrow-disk by the progenitor stellar wind.
On the other hand, the wide-disk model generates jet-like shapes like those displayed by the SNRs G290.1-0.8, S47, and G309.2-0.6.

\section*{Acknowledgements}
We appreciate the constructive comments of the referee, which allowed us to improve the previous version of this manuscript substantially.
PFV, AEC-A, JCT-R, and AE acknowledges financial support from PAPIIT-UNAM grants IG100422, IN113522, and IA103121. EMR and EMS are members of the Carrera del Investigador Cient\'\i fico of CO\-NI\-CET, Argentina. EMR is partially funded by grant PIP 112-201701-00604CO. AC specially acknowledges Triantafyllakia Institute for the support and hospitality. We thank Enrique Palacios (ICN-UNAM) for maintaining the Linux cluster, where the simulations were
performed.
The authors acknowledge the North-German Supercomputing Alliance (HLRN) for providing HPC 
resources that have contributed to the research results reported in this paper. 
PFV dedicates this work to the memory of professor Silvia Duhau (Buenos Aires University), who recently passed away.

\section*{Data Availability}
The data underlying this article will be shared on reasonable request to the corresponding author.



\bibliographystyle{mnras}
\bibliography{ref} 








\bsp	
\label{lastpage}
\end{document}